\documentclass[prl,superscriptaddress,twocolumn,showpacs]{revtex4}

\usepackage{amsmath}
\usepackage{amssymb}
\usepackage{bm}
\usepackage{epsfig}
\usepackage{graphicx}
\usepackage{color}

\renewcommand{\Re}{\mathop{\rm Re}}

\newcount\timehh  \newcount\timemm
\timehh=\time \divide\timehh by 60
\timemm=\time
\begin{document}

\title{Theory of spin noise in nanowires}
\author{M. M. Glazov}
\affiliation{Ioffe Physical-Technical Institute RAS, 194021
  St.-Petersburg, Russia}
\email{glazov@coherent.ioffe.ru}
\author{E. Ya. Sherman}
\affiliation{Department of Physical Chemistry, The University of the
  Basque Country, 48080 Bilbao, Spain} 
\affiliation{IKERBASQUE Basque Foundation for Science, Bilbao, 48011
  Bizkaia, Spain} 


\begin{abstract}
We develop a theory of spin noise in semiconductor nanowires
considered as prospective elements for spintronics.
In these structures spin-orbit coupling can be realized
as a random function of coordinate 
correlated on the spatial scale of the order of 10 nm. By analyzing different
regimes of electron transport and spin dynamics, we demonstrate that
the spin relaxation can be very slow and the resulting noise power
spectrum increases algebraically as frequency goes to zero. 
This effect makes spin effects in nanowires best
suitable for studies by rapidly developing spin-noise spectroscopy.
\end{abstract}


\maketitle


Nanostructures are the promising hardware elements for spintronics \cite{zutic:323} --
a rapidly developing branch of physics and technology aiming at studies
and application of spin-dependent phenomena in the charge transport
and information processing. The quest for the 
systems with ultralong spin relaxation times \cite{Wu201061} is one of the main challenges 
in this field. Since the dynamical spin fluctuations \cite{Ivchenko74} characterized
by correlations on the spin relaxation timescale, are seen as a spin
noise in the frequency domain, this search can be done with recently
developed highly accurate low-frequency spin noise spectroscopy ~\cite{Mueller2010} 
aimed at the measurement of intrinsic equilibrium spin
dynamics. The spin noise
spectroscopy allows to study the slow spin dynamics in (110)-grown quantum
wells~\cite{muller:206601} and in quantum dots~\cite{crooker2010}.
Theoretical background of this method is given, e.g., in Refs.~\cite{braun2007,starosielec:051116,kos2010}.

An interesting class of semiconductor nanostructures demonstrating peculiar and slow spin dynamics are the 
quantum wires~\cite{kiselev00,pramanik03,holleitner06}, where e.g. InAs, InSb as well as GaAs/AlGaAs systems 
are the prospective realizations. The effects of spin-orbit (SO) coupling on the
transport were clearly demonstrated there \cite{quay2010,Pershin04} and the nanowire based qubits were introduced \cite{Nadj2010,Bringer2011}. 
A SO coupling induced effective magnetic field acting on electron spins 
in nanowires is directed parallel or antiparallel to a certain axis~\cite{nishimura99,governale02,silva03,entin04,glazov04a} 
resulting in a giant spin relaxation anisotropy similar to that expected in some two-dimensional systems~\cite{averkiev:15582}. 
Since the SO coupling is a structure- and material-dependent property, 
all sorts of disorder (random doping~\cite{perel76eng,sherman03b,sherman03a,dugaev09}, interface fluctuations~\cite{golub04}, 
random variations in the shape, etc.) which cause electron scattering and nonzero resistivity,
can cause local variations in the coupling. As a result, 
in addition to the regular SO coupling, caused by the lack of bulk (Dresselhaus term) or structure 
(Rashba term) inversion symmetry, 
all low-dimensional structures inevitably have the random contribution in it. 
The spatial scale of the fluctuations is of the order of 10 nm as determined by the characteristic distances in
nanostructure were shown to give rise to a number of fascinating 
phenomena \cite{glazov05,GlazovSherman_rev,Strom}.
However, their role in quantum wires was not studied so far.

Here we address theoretically the electron spin dynamics in ballistic and
diffusive semiconductor nanowires aiming at the study of the spin noise
spectrum. Different regimes of electron spin relaxation are determined
and the crossovers between them are analyzed in detail. In particular, we
demonstrate that when the electron motion is diffusive and the dominant
contribution to the SO interaction is random, the spin relaxation becomes
algebraic rather than exponential and the spin noise power spectrum diverges
at low frequencies $\omega$ as $1/\omega ^{1/2}$, showing colored
noise \cite{Dutta1981,Weissman1988,Levinshtein1997} well suited for the
studies by the spin noise spectroscopy. A very slow spin
dynamics resulting in the low-frequency noise divergence  
makes nanowires an exception among semiconductor systems.

\begin{figure}[t]
\includegraphics[width=0.65\linewidth]{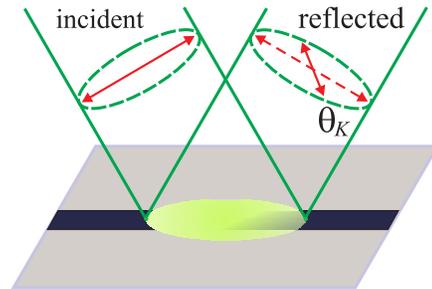}
\caption{Schematic plot of the experimental configuration: a quantum
wire (dark stripe) is illuminated by a linearly polarized beam 
and Kerr rotation angle of its
polarization plane $\theta_{K}$ is measured. Polarizations of the beams
are marked by  double-headed arrows. Dashed arrow corresponds to the polarization
of the reflected beam in the absence of the Kerr effect.}
\label{fig:kerr}
\end{figure}

The spin noise spectroscopy, 
reviewed in Ref.~\cite{Mueller2010},
is based on the optical monitoring of the spin fluctuations \cite{aleksandrov81} in 
Faraday, Kerr or ellipticity signals measured with a weak linearly polarized probe beam incident on
a single wire or a wire array sample, see Fig.~\ref{fig:kerr}. It can be shown similarly to
Refs.~\cite{kos2010,Mueller2010,zhu07} that for the probe 
tuned to the fundamental absorption edge, the Kerr rotation 
angle $\theta_K \propto s_z$~\cite{ppwires}, hence its
autocorrelation function is directly related to the spin noise:
$\langle \theta_K(t)\theta_K(t') \rangle \propto \langle s_z(t)s_z(t')\rangle$,  where 
$s_{z}(t)$ is the density of the $z-$component of the total electron spin. 
As a result, this optical technique measures
long-range correlations of equilibrium spin fluctuations occurring in the illuminated spot. 


We consider a single channel quantum wire extended along the $x-$axis
and represent the SO 
Hamiltonian as:
\begin{equation}
\label{HSO}
\mathcal H_{SO} = \frac{1}{2}[\alpha(x)  k_x+ k_x \alpha(x)]\sigma_{\lambda}.
\end{equation}
Here $k_x = -\mathrm i \partial/\partial x$ is the electron wave vector component along the wire axis, $\alpha(x)$ is the coordinate-dependent
SO coupling strength. In Eq.~\eqref{HSO} we assumed that the spin quantization axis, $\lambda$, is fixed, and 
$\sigma_\lambda$ is the component of spin operator along this axis. The specific form of the SO Hamiltonian Eq.~\eqref{HSO} implies that the 
effective field acting on electron spin points either parallel or antiparallel to the axis $\lambda$. This is obvious 
for a constant $\alpha(x)$ ~\cite{nishimura99,governale02,silva03,entin04,glazov04a}, and holds true provided that 
the microscopic symmetry of the fluctuations forming  the SO coupling randomness 
is the same as overall symmetry of the system.

The SO coupling is assumed to be the 
sum of the coordinate-independent contribution, $\alpha_0$, 
and the Gaussian random function with zero average, $\alpha_{\rm r}(x)$ such as 
$\alpha(x) = \alpha_0+\alpha_{\rm r}(x)$
with the correlation function ~\cite{GlazovSherman_rev}:
\begin{equation}
\label{Fcorr}
\langle\alpha_{\rm r}(x)\alpha_{\rm r}(x')\rangle = \langle \alpha_{\rm r}^2 \rangle F_{\rm corr}(x-x'),
\end{equation}
where $\langle \alpha_{\rm r}^2 \rangle$ is the mean square of SO coupling fluctuations and 
the range function $F_{\rm corr}(x-x')$.
We introduce also the typical correlation length of the SO coupling
\begin{equation}
\label{ld}
l_d = \int_0^{\infty} F_{\rm corr}(x) \mathrm d x,
\end{equation}
characterizing the size of the correlated domain of the random SO coupling.
Details of the models of random SO coupling can be found in Ref.\cite{GlazovSherman_rev}.

We begin with the semiclassical regime, where SO coupling disorder is smooth on the scale of electron wavelength, 
$l_d \gg \lambda_{\rm F}$, where the wavelength of the Fermi level electrons 
$\lambda_{\rm F}=2\pi/k_{\rm F}$, with $k_{\rm F}$ being the Fermi wave vector for 
the degenerate electron gas. The Hamiltonian~\eqref{HSO} implies that the spin rotation 
angle around the $\lambda$-axis during the motion from the point $x_0$
to 
$x_1$ is
\begin{equation}
\label{theta}
\theta(x_1,x_0) = \frac{2m}{\hbar^2} \int_{x_0}^{x_1} \alpha(x') \mathrm dx',
\end{equation}
where $m$ is the electron effective mass.
Eq.~\eqref{theta} shows that the angle is solely determined by electron initial and final positions
and does not depend on the history of the motion between these points. 
This result, being well established for the systems 
with regular SO coupling~\cite{levitov03,entin04,Tokatly20101104,Tokatly2010,Pershin11} holds also for the nanowires with the SO coupling disorder. 
As it follows from Eq.~\eqref{HSO} the spin precession rate is proportional to the electron velocity and given coordinate-dependent function. 
Hence, it does not matter whether the electron starting from the point $x_0$ reached the point $x_1$ ballistically or diffusively: all 
contributions to spin precession of the closed paths, where electron passes the same configuration of $\alpha(x)$ in the opposite directions,  cancel
each other.

The temporal evolution of electron spin is directly related the
electron motion along the wire.  
We consider here spin projections at given $z$ axis, 
perpendicular to the spin quantization axis $\lambda$. 
Time dependence of
electron spin $z$ component averaged over its random spatial motion and over
the random precession caused by the field $\alpha(x)$ can be
most conveniently characterized by the correlator 
$\langle 
s_{z}(t)s_z(0)\rangle =\langle s_{z}^2(0)\rangle \mathcal{C}_{ss}(t)$ with the
normalized correlation function: 
\begin{equation}
\label{sz:gen}
\mathcal{C}_{ss}(t) = \int_{-\infty}^\infty \mathrm d x\ p(x,t) \langle \cos{[\theta(x,0)]} \rangle,
\end{equation}
where $p(x,t)$ is the probability that electron travels distance $x$
during the time $t$. Note, that $\mathcal C_{ss}(t)$ can be
understood as disorder-averaged electron spin $z$ component 
found with the initial condition $s_z(0)=1$. It results from the
linearity of the spin dynamics equations: the correlators $\langle
s_i(t)s_j(0)\rangle$ satisfy exactly the same equations as average
values $\langle s_i(t)\rangle$ ($i,j=x,y,z$). In derivation of
Eq.~\eqref{sz:gen} we   
assumed also that the scattering of electrons, which
determines $p(x,t)$ is 
not correlated with the  
random SO field $\alpha_{\rm r}(x)$, hence, the averaging over the 
realizations of $\alpha_{\rm r}(x)$ denoted by the angular brackets and over the 
trajectories can be considered
independently. This can occur in 
nanowires where random Rashba fields are induced by doping while the
momentum scattering is due to the wire width fluctuations. 
If the same local
disorder determines the electron scattering and random SO fields, in
relatively clean systems the electron mean free path $l$ exceeds by far the
disorder correlation length $l_{d}$ in Eq.\eqref{ld}. Hence, spatial
scales of two 
random processes: $l$ for the electron backward scattering in the random
potential and $l_{d}$ for the spin precession are strongly different. As a
result, on the $l$-scale, the memory of the short-range correlations is
lost, and Eq.~\eqref{sz:gen} holds. Although Eq.~\eqref{sz:gen} is presented
for the smooth SO coupling disorder, where the electron motion is
semiclassical, $l_{d}/\lambda _{\mathrm{F}}\gg 1$, a general Green's
function approach confirms it for arbitrary $l_{d}/\lambda_{\mathrm{F}}$ values.

Our next step is to perform averaging of $\cos {[\theta (x,0)]}$ in
Eq.~\eqref{sz:gen} 
over the random realizations of the $\alpha (x)$-field. For this purpose 
we recast
\begin{equation}
 \cos{[\theta(x,0)]} = \Re{ \left\{ \exp{\left(\mathrm i \frac{2m \alpha_0}{\hbar^2} x\right)} 
\exp{\left[\mathrm i \vartheta_{\rm r}(x) \right]}\right\} },
\end{equation}
where 
\begin{equation}
\vartheta_{\rm r}(x)= 2m/\hbar^2 \int_{0}^{x} \alpha_{\rm r} (x') \mathrm dx'
\end{equation}
is the contribution of the random SO coupling into the spin rotation angle.
We expand last exponent in series in $\vartheta_{\rm r}$ assuming the Gaussian 
SO coupling disorder. In the averaging, odd powers of spin rotation angle vanish, $\langle \theta_{\rm r}^{2n+1}(x) \rangle =0$, 
for integer $n$ and even powers can be expressed solely with $\langle \theta_{\rm r}^2(x)\rangle$ as
\begin{equation}
\langle \theta_{\rm r}^{2n}(x) \rangle = \left \langle \left[  \frac{2m}{\hbar^2} \int_{0}^{x} \alpha_{\rm r}(x') \mathrm dx' \right]^{2n} \right \rangle 
= (2n-1)!! \langle \theta_{\rm r}^2(x)\rangle^n.
\end{equation}
Direct calculation shows that the mean square $\langle\theta_{\rm r}^2(x)\rangle$ caused by the random SO interaction is given by
\begin{equation}
\label{theta2}
\langle \theta_{\rm r}^2(x) \rangle  = 2 \left(\frac{2m}{\hbar^2}\right)^{2} \langle \alpha_{\rm r}^2 \rangle \int_0^x \mathrm d x' \int_0^{x'} \mathrm dy F_{\rm corr}(y).
\end{equation}
Finally, Eq.~\eqref{sz:gen} reduces to
\begin{equation}
\label{sz}
\mathcal{C}_{ss}(t) = \int_{-\infty}^\infty \mathrm d x\ p(x,t) \cos{\left(\frac{2m\alpha_0}{\hbar^2}x\right)}  \exp{\left[- \langle \theta_{\rm r}^2(x) \rangle/2\right]}.
\end{equation}
When $\langle\theta_{\rm r}^2(x)\rangle$ becomes considerably larger than one, spins are completely dephased. 
Equation~\eqref{sz} is our central result: it relates temporal average spin dynamics with electron motion
along the wire. Distribution function of electron 
displacements, $p(x,t)$, presented for different regimes of 
electron motion below, enables us to calculate spin 
evolution by Eq.~\eqref{sz}. The spin noise power spectrum is given by 
the transform of $\mathcal{C}_{ss}(t)$~\cite{kos2010}:
\begin{equation}
\left\langle s_{z}^{2}\right\rangle_{\omega }=2\int_{0}^{\infty }\mathcal{C}%
_{ss}(t)\cos \left( \omega t\right) {\rm d}t.  \label{snoise}
\end{equation}

\begin{figure}[t]
\includegraphics[width=0.6\linewidth]{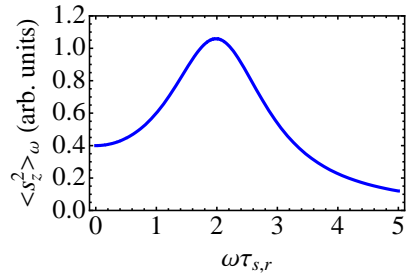}
\caption{Spin noise power spectrum, $\langle s_{z}^{2}\rangle_{\omega}$, for ballistic  
propagation, $\Omega_0\tau_{s,\rm r} =2$.
Due the exponential decay in Eq.\eqref{sz:bal} it is finite at $\omega=0$ 
with the width determined by the spin relaxation time $\tau_{s,\rm r}$. The spectrum peaks at the 
frequency $\Omega_0$ since average electron spin rotates in the SO field at the 
rate $\Omega_0$  and
asymptotically decays as $\omega^{-2}$ in accordance with the fluctuation-dissipation theorem.
}
\label{fig:snoise:1}
\end{figure}

To get a better insight into the problem, we begin with the key limits
\cite{footnote}. 
First, for the ballistic electron dynamics 
$p(x,t) = \delta(x-v_{\rm F} t)$, where $v_{\rm F} = \hbar k_{\rm F}/m$ is the 
Fermi velocity. The ballistic motion is realized on the temporal 
scale $t\ll \tau=l/v_{\rm F}$ with $\tau$ being the momentum relaxation time. 
We are interested in the spin dynamics on the time scale $t\gg \tau_d
= l_d/v_{\rm F}$, where $\tau_d$ is the time  
during which electron passes the correlated interval of the SO coupling 
fluctuations. Using Eq.~\eqref{sz} we obtain damped oscillations of
the spin $z$-component: 
\begin{equation}
\label{sz:bal}
\mathcal{C}_{ss}(t) 
 \approx \cos{\left(\Omega_0 t\right)} \exp{\left(-t/\tau_{s,{\rm r}} \right)},
\end{equation}
with the frequency $\Omega_0={2m\alpha_0 v_{\rm F}}/{\hbar^2}$ determined by the averaged 
SO coupling and the decay time caused by the SO coupling fluctuations
\begin{equation}
\label{tau:sd}
\frac{1}{\tau_{s,{\rm r}}} = \left(\frac{2m v_{\rm F}}{\hbar^2}\right)^{2} \langle \alpha_{\rm r}^2\rangle \tau_d.
\end{equation}
Equation~\eqref{tau:sd} for the spin relaxation time $\tau_{s,{\rm
    r}}$ is a result of 
random spin precession~\cite{GlazovSherman_rev}. Spin noise power spectrum 
calculated using Eqs.~\eqref{snoise} and \eqref{sz:bal} reads:
\begin{equation}
\label{s:noise:bal}
\left\langle s_{z}^{2}\right\rangle_{\omega}=2\tau_{s,\rm r}
\Re\frac{1-\mathrm i\omega\tau_{s,\rm r}}{\Omega_0^2\tau_{s,\rm r}^2 + (1-\mathrm i\omega \tau_{s,\rm r})^2 } 
\end{equation}
with the result presented in Fig.~\ref{fig:snoise:1}. 

\begin{figure}[t]
\includegraphics[width=0.65\linewidth]{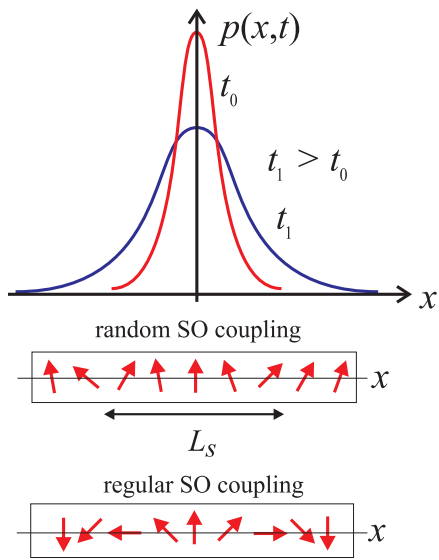}
\caption{Upper panel: Schematic illustration of the 
  displacements distribution $p(x,t)$ for two different time moments:
  $t_0<t_1$.  
Lower panels: The quantum wire and spins of diffusing electron for the
random and regular SO couplings, respectively. For the random SO coupling, 
if the electron is within the $L_s$ distance from its initial point 
[see Eq.~\eqref{sz:diff}] its spin is preserved, when it leaves this
interval, the spin dephases.}\label{fig:illustr}
\end{figure}

This ballistic regime of spin dynamics, however, can be realized only
in very clean systems, where $\Omega_0\tau \gg 1$.  
Otherwise, electron spin evolution occurs at the time scale, where
electron moves diffusively (Fig.\ref{fig:illustr}, upper panel), i.e.
\begin{equation}
\label{p:diff}
p(x,t) = \frac{1}{2\sqrt{\pi D t}} e^{-{x^2}/{4Dt}},
\end{equation}
where $D=v_{\rm F}^2\tau$ is the diffusion coefficient.
In the absence of the SO coupling fluctuations and provided 
that $\Omega_0\tau \ll 1$ exponential 
spin relaxation is due to the Dyakonov-Perel' mechanism~\cite{nishimura99,glazov04a} 
with the relaxation time $\tau_{s,\rm DP} = 1/(\Omega_0^2\tau)$. The
spin noise spectrum has a Lorentzian form 
$\langle s_{z}^{2}\rangle_{\omega}=2\tau_{s,\rm
  DP}/(1+\omega^{2}\tau_{s,\rm DP}^{2})$ with the width  
determined by the relaxation time. 

New physical features arise when the SO coupling fluctuations 
dominate over the regular contribution. 
From now on we put $\alpha_0=0$ and consider the system where SO
coupling is purely random 
and concentrate on the long-time ($t\gg \tau_d,\tau, \tau_{s,\rm r}$) 
dynamics. At these times, the system in Eq.~(\ref{sz}) is characterized by two
length parameters. One parameter is the diffusion length $\sqrt{Dt}$
in Eq.~(\ref{p:diff}), 
the other one   
\begin{equation}
L_{s} = \int_{0}^\infty \mathrm d x \exp{\left[- \langle \theta_{\rm r}^2(x) \rangle/2\right]},
\end{equation}
characterizes spin randomization. At sufficiently long times, when $\sqrt{Dt}\gg L_s$,  one
can take $p(0,t)$ instead of $p(x,t)$ and immediately obtain from Eq.~\eqref{sz} that the relaxation
is algebraic rather than exponential:
\begin{equation}
\label{sz:diff}
\mathcal{C}_{ss}(t)\approx p(0,t) \int_{-\infty}^\infty \mathrm d x \exp{\left[- \langle \theta_{\rm r}^2(x) \rangle/2\right]} = \frac{L_s}{\sqrt{\pi D t}}.
\end{equation}
Equation~\eqref{sz:diff} predicts extremely long spin decoherence described by the inverse square root 
law: $\langle s_z(t)\rangle \propto 1/\sqrt{t}$. This surprising result has a transparent physical 
interpretation (see Fig.~\ref{fig:illustr}): Indeed, if an electron is displaced from its initial position by a sufficiently large distance, $x\gtrsim L_s$, 
its spin rotation angle becomes so large, that it does not contribute to the total spin polarization 
owing to $\exp{\left[- \langle \theta_{\rm r}^2(x) \rangle/2\right]}$ in Eq.~\eqref{sz}. As a result, the 
spin polarization is supported by the electrons located in the vicinity of their initial positions, mainly 
due to the return after multiple scatterings by the random potential. 
The fraction of such electrons, in agreement with the diffusion distribution, decays as $p(0,t)\propto 1/\sqrt{t}$ 
resulting in the same behavior in the spin polarization. It is interesting to mention that this qualitative argument
does not work for the constant SO coupling despite spin of electron is restored upon the return to
the origin also here. The reason is that due the oscillations of the spin on the 
spatial scale of the order of $\hbar^2/m\alpha_0$ (see Fig.~\ref{fig:illustr}, lower panel) 
in Eq.(\ref{sz}), the 
diffusive return of electrons to the origin is insufficient for formation 
of the algebraic relaxation tail. 

Another realization of the $1/\sqrt{t}$ spin decay can be achieved for the very strong random SO 
couping where the spin relaxation occurs within one nanosize domain of the SO coupling, that
is at the electron displacement much less than $l_d$. 
In this case, spin relaxation rate is due to the Dyakonov-Perel' mechanism and is determined by 
the local value of $\alpha(x)$ inside the domain. Spins of electrons located in the intervals with large  
$\alpha(x)$ will relax fast, while spins of those experiencing weak $\alpha(x)$ will relax slow. 

Slow non-exponential spin relaxation, described by Eq.~\eqref{sz:diff}
manifests itself in the low frequency spin noise  
spectrum. From Eq.~(\ref{snoise}) it follows: $\left\langle
  s_{z}^{2}\right\rangle_{\omega} \propto {1}/{\sqrt{\omega}}$, 
i.e the spin noise diverges at $\omega\to 0$. Such a non-trivial
behavior is inherent  
to the quantum wires with random SO coupling, where spin restores upon return
to the origin: in multichannel wires for sufficiently fast
interchannel scattering~\cite{footnote1} and in two-dimensional systems spin 
relaxation is exponential~\cite{GlazovSherman_rev} and $\left\langle  s_{z}^{2}\right\rangle_{\omega=0}$ is finite.

To conclude, we studied theoretically spin noise in semiconductor nanowire
for different regimes of the electron propagation. We demonstrated that if
the spin relaxation is determined by the randomness in the SO coupling, spin
relaxation becomes algebraic being closely related to the high probability
for electron to stay close to its initial position as a result of a multiple
scatterings in the random potential. This behavior can appear in at least
two possible regimes: (i) when the electron motion is diffusive and (ii)
when the spin relaxation occurs on a small spatial scale of the order of 10
nm. In any of these cases, the spin noise power spectrum shows colored 
${1}/{\sqrt{\omega }}$ noise. In addition, this observation shows that
low-frequency optical spin noise spectroscopy is an excellent tool for
studying spin phenomena in semiconductor nanowires and characterization of
random potential and SO coupling there.

\textit{Acknowledgements} MMG is grateful to RFBR and ``Dynasty'' Foundation---ICFPM for financial support.
This work of EYS was supported by the University of Basque Country UPV/EHU grant GIU07/40,
MCI of Spain grant FIS2009-12773-C02-01, and "Grupos Consolidados UPV/EHU 
del Gobierno Vasco" grant IT-472-10.





\newpage

\widetext

\textbf{\centerline{Supplementary Material for ``Theory of Spin Noise in Nanowires''}}

\endwidetext

\section{SI. Spin noise at arbitrary frequencies}

Here we determine the spin noise spectrum for arbitrary
frequencies $\omega$. We employ the kinetic  
equation for electron distribution function $f(x,v_x,t)$ dependent on
the position, velocity $v_x$, and time. 
The equation has the form:
$$
 \frac{\partial f}{\partial t} + v_x \frac{\partial f}{\partial x}+
 \frac{f-\bar f}{\tau} =0, 
\eqno{\rm (S1)}
$$
where $f(x,v_x,t)$ satisfies the initial condition $f(x,v_x,0) =
\delta(x) [\delta_{v,v_{\rm F}} + \delta_{v,-v_{\rm F}}]/2$  
meaning that at $t=0$ it is built at $x=0$  with the equal fractions
of electrons 
with velocities $v_x= \pm v_{\rm F}$. As a result, the carriers can be
separated into  
the right movers, $v_x = v_{\rm F}$, and left movers, $v_x=-v_{\rm F}$ with function 
$\bar f = [f(x,v_{\rm F},t) - f(x,-v_{\rm F},t)]/2$ being the anisotropic
part of the distribution.  
The distribution of electron displacements is given by $p(x,t) =
f(x,v_{\rm F},t) + f(x,-v_{\rm F},t)$. 
It can be shown that the spatial Fourier transform and $\cos(\omega t)$ transform of
this distribution, $\tilde p(k,\omega)$ has the form
$$
\label{pko}
\tilde p(k,\omega) = 2\Re\frac{\tau(1-\mathrm i \omega\tau)}{(kl)^2
-\mathrm i \omega\tau(1-\mathrm i\omega\tau)}. 
\eqno{\rm (S2)}
$$
In accordance with Eq.~(11) in the main text the spin noise spectrum can be
presented as 
$$
\label{snoise1}
\left\langle s_{z}^{2}\right\rangle_{\omega} = 
\int_{-\infty}^\infty \frac{\mathrm d k}{2\pi}\ \tilde p(k,\omega) \mathcal T(k),
\eqno{\rm (S3)}
$$
where 
$$
\mathcal T(k) = \int_{-\infty}^{\infty}  \exp{\left[\mathrm i k x-
\langle \theta_{\rm r}^2(x) \rangle/2\right]} \mathrm dx.
\eqno{\rm (S4)}
$$

\begin{figure}[hptb]
\includegraphics[width=0.9\linewidth]{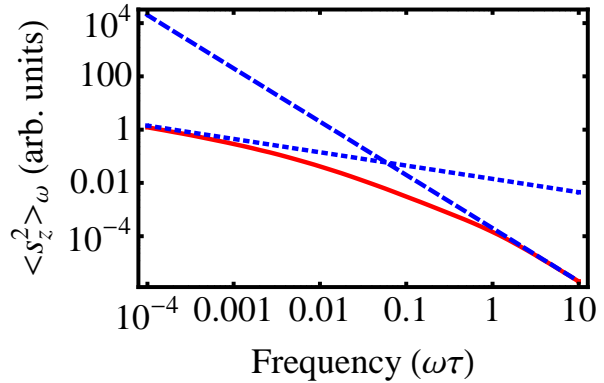}
\caption{Spin noise power spectrum for 
diffusive electron propagation, $\Omega_0\equiv 0$, 
$\Omega_{\rm r}\tau_d = 0.01$, $\tau_d/\tau=0.1$. 
Solid line shows exact result, calculated according to Eq.~(S6). 
Dotted (with the slope -1/2) and dashed (with the slope -2) lines show the low-frequency 
and high-frequency asymptotic, respectively.
}
\label{fig:snoise:full}
\end{figure}

Analytical result can be obtained in the regime where spin rotation
angles within each correlated domain of the SO coupling are small,
that is $\Omega_{\rm r} \tau_d \equiv
2m\sqrt{\langle\alpha_r^2\rangle}l_d/\hbar \ll 1$ 
with $\Omega_{\rm r}\equiv2\sqrt{\langle\alpha_r^2\rangle}k_{x}/\hbar$. 
Here the spin dynamics occurs on the spatial scale $x\gg l_d$,
mean squares of spin rotation angles are proportional to electron
displacement $\langle \theta_{\rm r}^2(x)\rangle\approx 2(\Omega_{\rm
  r}\tau_d)^2|x|/l_d$ being valid for $x\gg l_d$ or at $t\gg
\tau_d$, and function $\mathcal T(k)$ takes the form: 
$$
\label{tk}
\mathcal T(k) = \frac{2l_d (\Omega_r\tau_d)^2}{(\Omega_r\tau_d)^4+(kl_d)^2}.
\eqno{\rm (S5)}
$$
After lengthy transformations we obtain 
$$
  \label{snoise:analyt}
 \langle s_z^{2}\rangle_{\omega} = 2\Re
 \frac{\tau_d}{(\Omega_r\tau_d)^2  
\sqrt{{\mathrm i\omega\tau}/{(\mathrm i \omega\tau-1)}} -\mathrm i
\omega\tau_d} .
\eqno{\rm (S6)} 
$$
It can be seen from Eq.~(S6)
 that at low frequencies, 
$\omega\ll \tau_d^{-1},\tau^{-1}$, spin noise spectrum has the form:
$$
\left\langle s_{z}^{2}\right\rangle_{\omega} = 
\frac{\sqrt{2}\tau_{s,\rm r}}{\sqrt{\omega\tau}}, 
\eqno{\rm (S7)}
$$
in agreement with the analysis above. For high frequencies
$\omega\tau \gg 1$,   
$\left\langle s_{z}^{2}\right\rangle_{\omega}$ is given by
$2/(\omega^{2}\tau_{s,\rm r})$ since at $\tau_d \ll t \ll
\tau$ the electron motion is ballistic, and spin dephasing is caused by the random fluctuations of the
spin-orbit coupling, cf. Eq.~(12) of the main text.  
The entire 
frequency dependence of $\left\langle s_{z}^{2}\right\rangle_{\omega}$ is plotted
in Fig.~\ref{fig:snoise:full}.

\section{SII. Spin dynamics and noise in  
multichannel wires with random spin-orbit coupling}

The spin evolution in multichannel structures depends on the additional
set of parameters, $\{\tau_{i,j}\}$ being the scattering times between the channels $i$ and $j$,
as well as on the details of spin dynamics in every channel. 
For qualitative analysis (a general case requires a separate treatment) 
we consider a structure with two conducting channels,
where (i) spin-orbit coupling disorder in different channels is not
correlated and  
(ii) in each channel $\Omega_{\rm r}^{[c]}\tau_d \ll 1$ (superscripts denote channels), i.e. spin
rotation angles in correlated domains of the spin-orbit coupling are
always small. Here we can characterize the interchannel scattering by
a single time $\tau_{c}$ and 
focus on the most interesting case with no regular contribution to the
spin-orbit field: $\alpha_0\equiv 0$.

In the limit of very rare interchannel scattering events (the condition is
given below) the channels are independent. Hence, the general results expressed by 
Eqs.~(10), (11) of the main text and by Eq.~(S6) as well as asymptotic Eqs.~(S7) and (17) of the main
text hold. Although in these equations one has to average over
the realizations of $\alpha_{\rm r}^{[c]}(x)$ in different channels, 
the low-frequency spin noise power spectrum remains $1/\sqrt{\omega}$, 
the same as in a single channel wire.

Now we turn to the efficient interchannel scattering with short $\tau_c$. If $\tau_c \ll  \tau_d$ electron quits given channel faster than it quits
  the correlated domain. Spin rotations between interchannel
  scattering events are uncorrelated and, due to this randomness, spin dynamics is exponential:
$$
\langle s_z(t) s_z(0) \rangle \propto \exp{\left(-{\Gamma_{c}{t}}\right)},  \eqno{\rm (S8)}
$$
where the relaxation rate $\Gamma_{c}$ is of the order of 
$\left[  \max{ \left\{\Omega_{\rm r}^{[1]},\Omega_{\rm
      r}^{[2]} \right\}} \right]^2 \tau_c.$ 
Similar exponential decay of the spin correlator remains for $\tau_d
  \ll \tau_c \ll \tau$. Here, the mean square of the spin
  rotation angle between interchannel scatterings can be estimated as
  $\langle(\delta\Phi)^2\rangle = \langle \theta_{\rm r}^2(v^{[c]}\tau_c)\rangle \propto \tau_c$, with $v^{[c]}$
being the characteristic velocity in the channel. Spin relaxation is 
governed by the Dyakonov-Perel'-like mechanism, with the rate 
$$\Gamma \propto \langle (\delta \Phi)^2\rangle /\tau_c \sim
  \tau_{s,\rm r}^{-1}, \eqno{\rm (S9)}$$ 
which is $\tau_c$-independent for exactly the same reason as the spin relaxation rate due
to the random spin-orbit [Eq.(13) in the main text] does not explicitly depend on the electron free path.

Most interesting physics appears for a very weak interchannel scattering, $\tau_c \gg \tau,
\tau_d$. In this case, electron moves diffusively in a given channel
before the interchannel scattering occurs. As we have shown above, the
$1/\sqrt{t}$ tail in the spin polarization (and corresponding $1/\sqrt{\omega}$
spin noise) results from the carriers dwelling around the initial
point of their trajectories. Since the tail is formed at long times
$\sqrt{Dt} \gg L_s$, see Eq. (16) of the main text, it is supported by
electrons which moved many times back and forth in the random
potential. If the interchannel scattering is probable, electron may
return to the initial point via other channels, where its spin
rotation is not correlated with that in the initial one. Therefore, in
general $1/\sqrt{t}$ tail is destroyed and the usual exponential spin
relaxation takes place. However, if $\tau_c$ is long
enough to assure that the typical electron displacement
during the diffusion between interchannel scatterings
$L_c = \sqrt{D\tau_c} \gg L_s$, 
there is a time interval ${L_s^2}/{D}\ll t \ll\tau_c$ and the corresponding frequency range,
where the spin dynamics and the noise are algebraic:
$$
\langle s_z(t) s_z(0) \rangle \propto \frac{1}{\sqrt{t}}, \
\langle s_z^2\rangle_\omega \propto \frac{1}{\sqrt{\omega}}. \eqno{\rm (S10)}
$$
In the regimes of a highly efficient 
interchannel scattering, with the spin relaxation described by Eqs.(S8) or (S9), the probability
of spin components restoration upon return to the initial position is strongly suppressed, and, as a result,
spin noise power spectrum at low frequency decreases and becomes finite. 

For completeness, we mention that if the random spin-orbit coupling does not depend on the
channel, that is $\alpha_{\rm r}^{[1]}(x)=\alpha_{\rm r}^{[2]}(x)$, spin precession angle between any points
$x_1$ and $x_2$ is insensitive to the interchannel scattering, and our analysis in the main text holds exactly the same.

\end{document}